\documentclass{svjour3}                     
\usepackage{graphicx}
\usepackage{float}
\begin{document}

\title{Coverage threshold for laser-induced lithography}

\author{Weliton S. Martins \and Marcos Ori\'{a} \and Thierry Passerat de Silans \and Martine Chevrollier.}

\institute{Weliton S. Martins \and Marcos Ori\'{a} \and Martine Chevrollier \\ \email{martine.chevrollier@ufrpe.br}    \at
              Universidade Federal Rural de Pernambuco - Unidade Acad\^{e}mica do Cabo de Santo Agostinho
Rua Cento e Sessenta e Tr\^{e}s, 300 - Garapu - Cabo de Santo Agostinho - PE - 54518-430, Brazil \\
           \and 
                   T. Passerat de Silans \at
              Departamento de F\'{i}sica/CCEN, Universidade Federal da Para\'{i}ba, Caixa Postal 
5008, 58051-900, Jo\~{a}o Pessoa-PB, Brazil
}


\maketitle

\begin{abstract}
Recent experimental observations of laser-induced adsorption at the interface between an alkali vapor and a dielectric surface have demonstrated the possibility of growing metallic films of nanometric thickness on dielectric surfaces, with arbitrary shapes determined by the intensity profile of the light. The mechanisms directly responsible for the accumulation of atoms at the irradiated surface have been shown to involve photo-ionization of atoms very close to the surface. However, the existence of a vapor-pressure threshold for initiating the film growth still raises questions on the processes occurring at the surface. In this letter, we report on the observation that the vapor-pressure threshold corresponds to a minimum adatom coverage necessary for the surface to effectively neutralize the incoming ions and make possible the growth of a multilayer film. We discuss the hypothesis that the coverage threshold is a surface conductivity threshold.
\end{abstract}

\section{Introduction}
\label{intro}
The search for methods to grow nanometric films on a substrate is a modern technological demand. For a long time, lasers have been used in processes of lithography aiming at fabricating such nanometric-size features. UV lasers mostly, using masks or not, interact directly with the surface of the manufactured material \cite{ehrlich1989laser}. Direct manipulation of atoms and molecules with light is also possible. An important avenue has been opened in the control of the deposition of cold atoms on surfaces by the application of optical forces with lasers, that guide particles through optical patterns \cite{meschede2003atomic}. More recently, a new technique of control of the atomic binding to a surface was explored \cite{Afanasiev2007,Martins2013}, whereby the manipulation of the internal state of atoms very close to the surface is a key element of the process and that present the important feature of depositing a metallic film, with arbitrary shape and nanometric thickness resolution, directly from a thermal atomic vapor. This technique is the subject of the present study.\\
Optimizing the performance of this lithographic technique requires knowledge of the atom-surface interaction, and in particular of the adsorption process. Interactions between atoms and surfaces can be modified by light. For example, fundamental and technological aspects of light-induced atomic \textit{desorption} (LIAD) \cite{Meucci1994,Torralbo-Campo2015} and diffusion \cite{Vartanyan} have been thoroughly investigated in the past decades. Less intuitive, the phenomenon of light-induced \textit{adsorption} has been evidenced in 2007 \cite{Afanasiev2007} and the role of the laser in this process has been described \cite{Martins2013}. This laser-induced adsorption process allows for the growth, on a dielectric surface, of a metallic film of nanometric thickness whose shape is determined by the intensity profile of the light. Systematic measurements \cite{Martins2013} have shown that the film is grown from ions created by the interaction of the laser with the vapor and that, for a given surface temperature, a minimum vapor pressure is needed for the laser-induced film to start growing. In this letter, we investigate the origin of this vapor-pressure threshold, assuming it is related to the requisite that the ions be neutralized at the surface in order to suppress repulsion between them and allow for their accumulation as a metallic film. In section \ref{sec:experiment}, we describe the experimental set-up used in the investigation of the laser-induced adsorption mechanism. In section \ref{sec:results}, we recall the role of the laser in the light-induced adsorption process \cite{Martins2013} and we discuss the measurements that allow us to relate the observed vapor-pressure threshold to the surface state before exposure to the laser light. In section \ref{sec:adsorption} we review some of the literature findings on the properties of adsorption of alkali atoms on dielectric surfaces. In section \ref{sec:neutralization}, based on these findings, we substantiate our hypothesis on the origin of the vapor-pressure threshold for growth of a metallic film from Cs ions. We claim that the threshold originates from various surface mechanisms that converge to create a specific condition of surface conductivity, which in turn permits the neutralization of the ions. We conclude in section \ref{sec:conclusion}.

\section{Light-induced lithography Experiment} 
\label{sec:experiment}

\begin{figure}
\includegraphics[width=8cm]{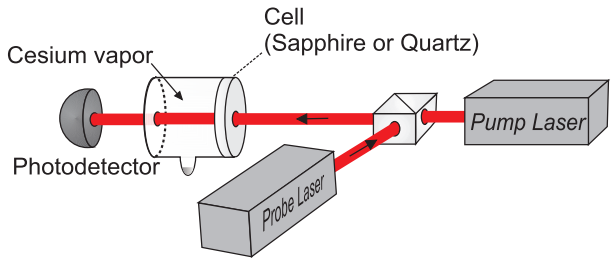}
\caption{\label{fig:setup} (Color online). Experimental setup to investigate light-induced adsorption. The film-inducing pump laser is resonant with the $D_2$ line of cesium and is incident at the interface between the cell window and the vapor. The pump beam is filtered by an optical fiber (not shown) in order to produce a Gaussian profile with radius of 1mm. The windows and reservoir of the 2 cm-long cell are independently heated with ovens (not shown). The film thickness is measured from the transmittance of a low-intensity HeNe probe laser.}
\end{figure}

The experimental apparatus that we use to observe light-induced adsorption phenomena is shown in Fig. \ref{fig:setup}. It consists of an optical cell containing a low-pressure cesium vapor, illuminated by a laser beam with frequency tuned inside the Doppler-broadened  $D_2$ line of cesium (Pump Laser in Fig. \ref{fig:setup}). Under particular conditions of vapor pressure and surface temperature, discussed below, cesium atoms from the vapor phase start accumulating at the illuminated region of the cell window, leading to the formation and growth of a metallic film on the inner surface of the illuminated window \cite{Afanasiev2007,Martins2013}. The thickness of the film reproduces the laser spatial profile \cite{Martins2013}. 
The thickness of the thin film on the inner surface of the cell window is probed by a weak laser beam, non-resonant with the vapor (He-Ne at 633 nm, Probe Laser in Fig. \ref{fig:setup}), whose transmission through the film is assumed to decay exponentially according to the Beer-Lambert law:

\begin{equation}
P = P_0 e^{-\beta L},
\label{eq:Beer}
\end{equation} 

where $P$ and $P_0$ are, respectively, the transmitted and incident laser power and $\beta$ is the absorption coefficient. We consider $\beta= 4 \pi n" / \lambda_{NR}$, with $\lambda_{NR}$ the wavelength of the non-resonant probe laser and $n"$ the imaginary part of the index of refraction of the film, taken here as the bulk extinction coefficient of cesium, $n"=1.28$ at 633 nm \cite{Palik1985}. In Eq. (\ref{eq:Beer}), $L$ is the average thickness of the adsorbate layer crossed by the laser beam. The diameter of the probe laser (1 mm) is smaller than the pump laser's (3 mm) so that we probe the central area of the Gaussian pump beam, where the intensity profile and thus the thickness are rather uniform. In the absence of pumping light, $L=L_{th}$ is the average thickness of the alkaline film deposited in both cell windows in thermal equilibrium with the atomic vapor. In the presence of pumping light, $L=L_{th}+ L_f$, where $L_f$ is the average thickness of the light-induced film. $L_f$ is the light-induced increment over the thermal film and is obtained by calculating the ratio between the power transmitted through the light-induced film at time $t>0$ and the power transmitted at dark thermal equilibrium (adlayer at time $t = 0$, when the pump laser is turned on) for the same surface and reservoir temperatures. 

The pump beam is blocked during the measurements of the probe beam. These measurements last 200 ms and are repeated every 5 s. \\

In both types of measurement (with or without film-inducing light), the intensity of the non-resonant probe beam (I$_{NR} \approx$ 0.6 mW/cm$^2$) is weak enough to avoid LIAD \cite{Meucci1994,Torralbo-Campo2015}. The vapor pressure in the cell body, of the order of tens of mTorr, is determined by the temperature of the Cs reservoir, which is controlled independently of the temperature of the cell windows. All temperatures are measured with an estimated precision of $1$ $^\circ$C. 

\section{Results}
\label{sec:results}

\subsection{Light-induced film} 
\begin{figure}
\centering
\includegraphics[width=8cm]{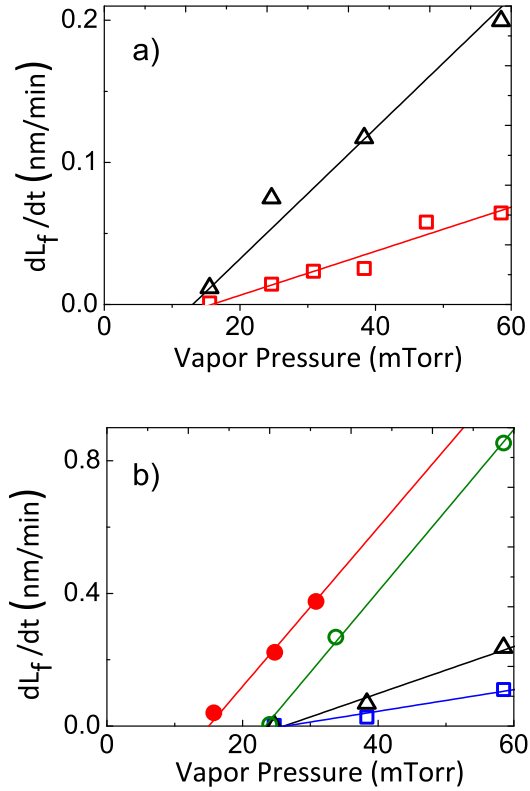}
\caption{(Color online). Growth rate of the light-induced film in the initial few-layer regime, as a function of the atomic vapor pressure in the cell, for various intensities ($I_P$) and frequency detunings ($\delta$) of the pump laser. (a) On a sapphire window, for $I_P$ = 3.2 mW/mm$^2$, $T_w$ = 215 $^\circ$C and two different frequency detunings: $\delta=$750 MHz (black triangles) and $\delta=$1200 MHz (red squares). (b) On a quartz window, for two  different window temperatures, $T_w$=190 $^\circ$C (solid symbols) and $T_w$=215 $^\circ$C (empty symbols):  $I_P$=4.4 mW/mm$^2$ and $\delta=1500$ MHz (circles), $I_P$=4.4 mW/mm$^2$ and $\delta=750$ MHz (triangles), $I_P$=2.5 mW/mm$^2$ and $\delta=750$ MHz (squares). The solid lines are guides to the eyes.}
\label{fig:density}
\end{figure}

We first recall the results obtained in \cite{Martins2013} from measurements of the growth rate of the thickness of the light-induced film, $L_{f}$. In the early stage of formation of the light-induced film, in which we are interested here, the growth rate of the film is constant whenever all the parameters are kept fixed. We have systematically measured the thickness growth rate as a function of the key parameters of the process (laser intensity \footnote{In the early stage of light-induced film growing, the thickness of the film is sufficiently small that the intensity of the pump laser can be considered constant across the film.} and frequency, window temperature and vapor pressure). From the dependence of the growth rate, cubic with the pump laser intensity and linear with the vapor pressure, we have inferred that the role of the pump beam is to ionize the cesium atoms in the vapor \cite{Martins2013}. At the high atomic densities where the formation of the film can actually be observed (see Appendix), the laser absorption by the vapor is very strong, resulting in a very small penetration of the beam. Therefore, the ions are produced very close to the window ($\leq$ 1 $\mu$m) and have a very high probability of interacting with the surface before leaving the thin illuminated volume. These ions constitute the raw material for the film \cite{Martins2013} and therefore have to be adsorbed at the surface in some way. However, in order to build up the observed multi-layer film with a thickness of a few nanometres, the net surface charge must be neutralized to overcome the mutual repulsive interaction between adsorbed and incoming ions. Here we investigate possible mechanisms for this neutralization of ions at the surface. The key feature that we explore for this purpose is the existence of a vapor-pressure threshold for the formation of the film. This threshold can be seen in Fig. \ref{fig:density}, where we plot the growth rate of the laser-induced film, measured as a function of the vapor pressure in various experimental conditions. Figure \ref{fig:density}(a) shows measurements carried out on a sapphire window. All other measurements reported in this article were made in a (fused-) quartz cell (Fig. \ref{fig:density}(b)). We see that variations of the intensity or frequency of the pumping light only affect the efficiency of the film growth, not the threshold value. This threshold depends on the temperature of the window (in Fig. \ref{fig:density} (b), two different temperatures result in two different values of the vapor-pressure threshold) and on the substrate material (Figs. \ref{fig:density}(a), sapphire and (b), quartz). We stress here that Figs. \ref{fig:density}(a) and (b) do not allow us to make quantitative comparisons between  the processes on sapphire and quartz surfaces, as the measurements were carried out in two cells differing in their thermodynamic characteristics and surface preparation.

The vapor-pressure threshold is established by the surface condition, i.e., by the thermal layer of adatoms (adlayer) present before exposure to the pump laser. These previously-adsorbed atoms modify the state of the surface and determine the ability of the surface to neutralize the incoming ions. As such, the threshold surface condition does not depend on the pump laser, as actually observed in Fig. \ref{fig:density}.\\
Further substantiation of this interpretation requires that we investigate the relationship between the vapor pressure and the state of the surface in thermal equilibrium, when no light is present. 

\subsection{Thermal equilibrium adlayer}
\label{sec:dyna }

\begin{figure}
\centering
\includegraphics[width=8cm]{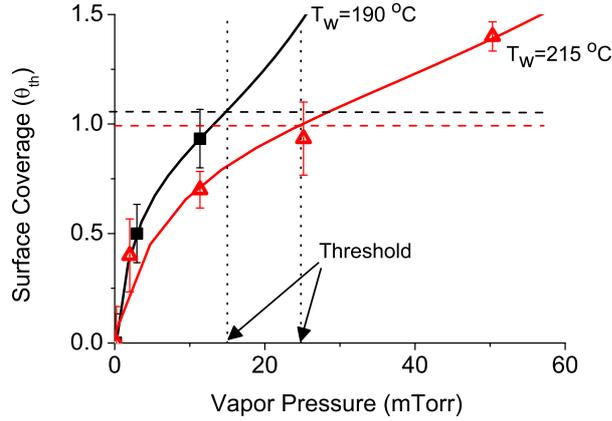}
\caption{(Color online) Surface coverage as a function of the Cs vapor pressure in the quartz cell in the dark regime, for the two temperatures of the window used in Fig. \ref{fig:density}(b). The solid lines are fits by a BET model (see Eq. (\ref{eq:BET})). Vertical dotted lines indicate the vapor-pressure thresholds for $T_W=190$ $^\circ$C and $T_W=215$ $^\circ$C obtained from Fig. \ref{fig:density}(b). The vapor pressure is determined from the reservoir temperature, with an estimated incertitude of 15\%.}
\label{fig:BET}
\end{figure}

The density of adsorbed atoms on the surface at thermal equilibrium is determined by detailed balance between adsorption and desorption processes. Following BET's model \cite{brunauer1938} for physisorption, when atom-surface attraction is due to dispersion forces, detailed balance is applied to each layer of a multi-layer film, neglecting lateral interaction between adatoms. For each layer $i$, the adsorption rate is proportional to the Cs vapor pressure $p$ and the desorption rate follows an Arrhenius law with binding energy $E_{A,i}$. We make the approximation that $E_{A,i}$ is equal to the heat of condensation, $E_c$, from the second layer on \cite{brunauer1938} ($E_{A,i}=E_C$ for $i\geq2$). Detailed balance between adsorption and desorption in each layer results in an equilibrium coverage of the surface, where the coverage is defined as the ratio between the total number of adsorbed atoms and the number of atoms in a monolayer \cite{brunauer1938}: 
\begin{equation}
\theta_{th} = \frac{cx}{1+(c-2)x-(c-1)x^2},
\label{eq:BET}
\end{equation} 
with $c=e^{\left(E_{A,1}-E_C\right)/k_BT_w}$ and $x=\frac{p(T_R)}{p(T_W)}$ the ratio between the vapor pressure in the cell (determined by the temperature of the cell's reservoir $T_R$) and the saturating vapor pressure at the surface temperature $T_w$. We take the average thickness of the cesium film on the surface, measured in the experiment, as proportional to the coverage, $L_{th}=L_M\theta_{th}$, where $L_M$ is the thickness of a complete monolayer.\\ 
We have measured the transmittance of the non-resonant probe beam as a function of the vapor pressure in the dark regime. At fixed surface temperature, the transmittance of the probe beam decays when the vapor pressure is increased, due to the increase of the flux of atoms toward the surface that results in the augmentation of the number of adsorbed atoms \footnote{We have verified that $P$ is only attenuated by surface adlayers. Cs dimer lines, for example, occur close to the wavelength of the He-Ne laser. Complementary experiments, carried out on a overheated cell where absorption of light by sparse adlayers is negligible (see section \ref{sec:dyna }), have shown no noticeable change of the transmitted power of the He-Ne laser as the vapor pressure was varied in the range explored in Figs. \ref{fig:density} and \ref{fig:BET}, thus excluding a possible contribution of absorption by impurities or dimers in the vapor. Since Cs dimers have also lines around 780nm, we have measured the transmittance of a 780nm laser beam through the vapor without noticeable modification when scanning the frequency around this wavelength.}. The measured surface coverage $\theta_{th}$ is shown in Fig. \ref{fig:BET} as a function of the vapor pressure for the two surface temperatures used in Fig \ref{fig:density}(b) for the quartz cell. The solid lines are fits to the experimental data using Eq. (\ref{eq:BET}). The adsorption energy obtained from the BET model is $E_{A,1}=E_C+(0.12\pm 0.03)$ eV.   Figure \ref{fig:BET} shows that the vapor-pressure thresholds for laser-induced film formation at $T_W=190$ $^\circ$C and $T_W=215$ $^\circ$ C (obtained from Fig. \ref{fig:density}(b) and indicated by vertical dotted lines) correspond to the same surface coverage. This supports the hypothesis that the vapor-pressure threshold is a coverage threshold. Note that at threshold the surface coverage approaches a complete monolayer, seen as an inflection point in the BET curve \cite{Adamson1997} in Fig. \ref{fig:BET}. The BET model gives $E_{A,1}\approx 0.89$ eV (for $E_C=0.75$ eV \cite{Kamins1968}), which is larger than values reported in the literature (for instance, $E_{A,1}=0.66$ eV in \cite{bouchiat1999} and $E_{A,1}=0.2$ eV in \cite{Freitas2002}). We believe that the reason for such a discrepancy is the over simplification of the BET model, that does not take into account, for instance, the adatom-adatom interaction \cite{BonchBruevich1997}. In the next section we describe the adsorption of Cs atoms on dielectric surfaces and discuss possible neutralization mechanisms.

\section{Adsorption of Cs atoms on dielectric surfaces} 
\label{sec:adsorption}

The adsorption of Cs atoms on dielectric surfaces schematically occurs according to one of the two following regimes: i) physisorption or ii) chemisorption. 
Weak dipole-dipole interactions are at the origin of physisorption phenomena. The atoms are then trapped in the potential well resulting from the long-range van der Waals attraction and the short-range Pauli repulsion between atomic and surface electrons. The depth of such a potential well ranges from a few meV to a few tenths of an eV and its minimum is localized at a fraction of a nanometer from the surface \cite{Hoinkes1980,desjonqueres2012}. There is no barrier to prevent the atom approaching the surface from entering the physisorption well and the kinetics of physisorption are fast. Physisorbed atoms retain most of their electronic structure and the valence electron remains attached to the atom's core \cite{Brause1997}. 
In chemisorption, on the other hand, the potential well is deeper (a few eV) and the minimum is closer to the surface than in the physisorption case \cite{desjonqueres2012}. The electronic structure of the atom-surface compound is strongly remodeled, with creation of a chemical bond between them. The strong binding between atom and surface gives rise to a partial charge transfer \cite{Lopez1999}. The binding energies in this case are typically on the order of a few eV and an activation barrier may be present. The adsorption energy strongly depends on the chemical properties of the adatom and of the surface site where the atom is adsorbed.\\

Some experimental \cite{Brause1997} and theoretical \cite{Lopez1999} works have investigated the details of the process of adsorption of cesium atoms on quartz (silica, SiO$_2$). The first atoms incident on a clean quartz surface are chemisorbed, mainly on defects corresponding to a nonbridging oxygen center (Si-O*) at the surface or, to a lesser extent, on defects corresponding to a missing oxygen on the surface (E' defect center, ≡Si*) \cite{Lopez1999}. 
The strong binding between the chemisorbed atom and the surface corresponds to a significant charge transfer where the Cs atom retains a net positive charge. This charge transfer induces an electric double layer at the surface that results in a decrease in the work function \cite{Langmuir1923,Topping1927,Gurney1935} up to about 3.5 eV in relation to the clean surface. This minimum work function is reached at low surface coverage \cite{Brause1997}, of the order of $10^{-2}$ . 
For higher coverages, the defect sites are saturated and adsorption occurs either at regular surface sites or around Cs atoms chemisorbed at defect sites, forming metallic clusters. In this second stage of adsorption, atoms are more weakly bound to the surface (physisorption) and the work function increases slightly in relation to its minimum, toward the value for the bulk adsorbate (2.1 eV for Cs \cite{fomenko2012}). The same behavior occurs for other alkali atoms on a variety of substrates \cite{gerlach1970,lang1971,vanWunnick1983,wilde1999}.\\
The coverage threshold for growing a light-induced film, as observed in our experiments, manifests itself in this regime where all the sites available for chemisorption are occupied and the work function of the substrate saturates to a value close to the work function of the bulk Cs. Furthermore, in the range of surface temperatures that we explore, the desorption of chemisorbed atoms is negligible. As a result, the number of chemisorbed atoms is constant and the work function of the surface is unchanged. Although taking place in a stationary regime of chemisorption (saturated chemisorption sites and work function), the existence of a threshold for film formation hints to a sudden ability of the surface to neutralize ions created in the vapor by the laser. This threshold depends on the coverage by physisorbed atoms, that we manipulate in our experiment.   

\section{Surface condition at the vapor pressure threshold: Possible neutralization scheme} \label{sec:neutralization}
When the substrate work function $W$ is lower than the atom energy of ionization, an electron can be transferred from the substrate to an ion, thus neutralizing it \cite{winter2002collisions,Martins2013}.
$W$ switches from larger to smaller than the Cs atom energy of ionization $I$ around a coverage of the order of $10^{-2}$, which is much lower than the threshold coverage $\theta_{th} \approx$ 1 measured in this work. In other words, although the condition $W \leq I$ is a necessary one for electrons to be preferentially localized at the atomic core, it is not sufficient for the effective neutralization of the incoming ions. We claim that the sought-after additional condition is a conductivity threshold, that allows for the displacement of electrons of the cesiated surface to landing sites of Cs ions on the surface.

\begin{figure}
\includegraphics[width=8cm]{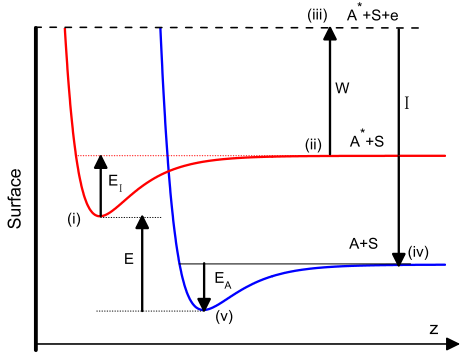}
\caption{(Color online). Typical surface potential wells for chemisorption, asymptotically yielding an ion free from the surface interaction (\textbf{A$^*$+S}), and for physisorption, asymptotically yielding a neutral free atom (\textbf{A+S}). We draw an hypothetical cycle from state (i) to state (v) that we use to determine the energy difference $E$ between chemisorption and physisorption states. In the configuration shown here, $E>0$, making the transfer from the chemisorption to the physisorption well highly probable. The picture is drawn using the energies $E_I=1$ eV, $W=2.1$ eV, $I=3.9$ eV and $E_A=0.75$ eV, corresponding to the values of our experimental parameters (see text).}
\label{fig:FigE}
\end{figure}

To determine whether the atom will preferentially adsorb as an ion or as a neutral atom, the direct comparison between the surface work function and the atomic ionization energy is an oversimplification since it does not take into account physisorption and chemisorption binding energies. Indeed, electrons can be transferred from the substrate surface to the adatom core by thermal excitation, leading the (atom+surface) system from a chemisorption to a physisorption state. The ratio, $R$, between the number of physisorbed atoms and the number of chemisorbed ions depends on the surface temperature and on the energy difference, $E=E_{chem}-E_{phys}$, between those two states \cite{Rasor1964}: $R\propto \textrm{exp}(E/k_BT_w)$. To determine $E$, we consider the hypothetical cycle shown in Fig. \ref{fig:FigE}, where two typical surface potential wells are shown: a chemisorption one, asymptotically yielding an ion free from interactions with the surface (\textbf{A$^*$}+S), and a physisorption one, asymptotically yielding a free neutral atom (\textbf{A+S}). We write the energy balance as the system follows an hypothetical cycle, e.g. from and to the chemisorption state (i), that is, the ion (\textbf{A$^*$}) trapped in the interaction potential well with binding energy $E_I$ and the surface (\textbf{S}) with electrons in its valence band. Energy $E_I$ being provided to the system, the ion is desorbed and the system evolves to a state (ii) with noninteracting ion and surface (\textbf{A$^*$+S}). Supply of an energy $W$, corresponding to the surface work function, removes an electron from the surface and converts the system into (state (iii)) a free ion, a free electron and the surface (\textbf{A$^*$+S+e}).
Then, the electron can bind to the free atom and release the energy corresponding to the atomic ionization energy (\textbf{I}), taking the system to state (iv), consisting of a free neutral atom and the surface (\textbf{A+S}). Trapping of the atom in the physisorption well (v) releases the energy $E_A$. Finally, the cycle is completed by providing the energy difference $E$ to the system, such that $E$ can be determined from the following relation \cite{Rasor1964}: 
\begin{equation}
E - E_A - I + E_I + W = 0.
\label{eq:Ebalance}
\end{equation}

We can estimate $E$ by using known values of the energies involved in the hypothetical cycle detailed above. At the threshold for growth of the light-induced film, the defect sites where chemisorption occurs are saturated and the surface work function approaches the Cs bulk value of 2.1 eV \cite{fomenko2012,Brause1997}. The chemisorption binding energy ranges from 1 eV  to 3 eV, with the lower binding energy at Si sp$^3$ dangling bonds and the higher binding energy at nonbridging oxygens defects \cite{Lopez1999}. We take the physisorption binding energy as being the heat of evaporation of bulk Cs, $E_C=$0.75 eV \cite{Kamins1968} (see section \ref{sec:dyna } above) and the atomic ionization energy as $I=$3.9 eV \cite{Weber1987}. The energy difference $E$ estimated from those values ranges from 1.55 eV (for $E_I=1$ eV) to -0.45 eV (for $E_I=3$ eV). Ad-ions may be easily neutralized around sites where $E$ has the largest, positive value (Si sp$^3$ dangling bonds on SiO$_2$ surface). The energies involved in the adsorption/desorption/neutralization process depicted in Fig. \ref{fig:FigE} are subject to variations. For example, due to the attractive interaction between adsorbed atoms, the physisorption potential depth $E_A$ itself changes with the number of adsorbed atoms \cite{Kamins1968}. The physisorption energy considered, 0.75 eV, corresponding to $E_C$, is expected for atoms in the outside layer of the adsorbed film \cite{Kamins1968}, since they bind to other Cs atoms. The physisorption energy of inner adatoms in a cluster is expected to be larger than 0.75 eV, possibly rendering the energy difference $E$ positive even for $E_I$ = 3 eV, and, thus, favoring neutralization of the arriving ad-ions. However crude the above estimate of $E$ may be, it gives good insight into the surface predisposition to transfer the incoming Cs ions from an ionic bond to a neutral one, opening the way to their accumulation as a film. Incoming ions are neutralized if this surface state condition is met simultaneously with the availability of electrons. \\

Adsorption of alkali atoms on dielectric surfaces has the effect of significantly rising the surface conductivity \cite{Papoyan2002,Agnew1995,bouchiat1999}. The main feature behind this increase of surface conductivity is the creation of metallic paths through which the electrons can flow. This situation occurs when the clusters are large enough to merge. Before that, the conductivity is already increased by the possibility of electron tunnelling between clusters \cite{morris1977}. The surface conductivity should therefore increase with the number of physisorbed atoms. In our experiment, it is thus expected that higher Cs vapor pressures should yield a higher surface conductivity. At the same time, the existence of atoms in physisorption states results in the appearance at the surface of a band structure at the energy of the Cs $6S_{1/2}$ level \cite{Brause1997}. The rise of the surface conductivity allows the electrons available in this new surface band structure to diffuse on the surface, eventually neutralizing incoming ions. The coverage threshold for ad-ion neutralization should therefore correspond to a threshold for surface conductivity. As discussed in the previous paragraph, the diffusing electrons that interact with an incoming ion encounter a favorable energetic condition to be preferentially transferred to the adatom core, thus fulfilling the ion neutralization requirement for the growth of a multilayer film.\\

\section{Conclusion} 
\label{sec:conclusion} 
We have investigated the origin of the vapor-pressure threshold for the light-induced adsorption of an alkaline vapor on a dielectric surface. The role of the light in this adsorption mechanism is to ionize atoms very close to the surface ($\leq 1\mu$m). The ions produced stick to the surface and their net charge must be neutralized for a multilayer film to grow. The neutralization mechanism occurs when a specific surface condition, corresponding to the vapor-pressure threshold, is met. By measuring the thickness of the metallic film as a function of vapor pressure in the (dark) thermal regime, we have inferred that the vapor-pressure threshold is an adatom coverage threshold. The transfer from a chemisorbed ad-ion state to an physisorbed adatom state is possible if the energy difference between those states is of the order of the thermal energy. The initial decrease of the work function by thermally chemisorbed atoms makes this condition fulfilled at some surface defects and on regular sites. The increase of the number of physisorbed atoms increases the surface conductivity by creating metallic paths on the surface. It also results in the appearance of a new surface band structure around the energy of the $6S$ atomic level of Cs. We believe that the rise of the surface conductivity favors the electrons of this metallized band to diffuse on the surface and to recombine with the arriving photo-induced ion. The positive energy difference between chemisorption and physisorption states ($E_{chem}> E_{phys}$) favors the transfer of the surface electron to the adatom core. By the picture we draw here the experimentally measured surface coverage threshold for the growth of the laser-induced film is a surface conductivity threshold. Further experimental and theoretical investigations are needed to characterize this process, as well as to determine the properties and quality of the resulting film.

\begin{acknowledgements}
The authors acknowledge financial support from Brazilian funding agencies: Conselho Nacional de Desenvolvimento
Cient\'{i}fico e Tecnol\'{o}gico (CNPq), Coordena\c{c}\~{a}o de Aperfei\c{c}oamento de Pessoal de N\'{i}vel Superior (CAPES), and Financiadora
de Estudos e Projetos (FINEP).
\end{acknowledgements}

\bibliographystyle{spphys}       
\bibliography{adsorption}
\newpage

\section{Appendix:Vapor-pressure threshold in a quartz cell with Cs vapor}

\begin{figure}[H]

The existence of the vapor-pressure threshold for growing a light-induced film prevents this phenomenon to be commonly observed in atomic physics experiments using optical vapor cells. Indeed, it is usual to rise the windows temperature relative to the reservoir to prevent Cs condensation on the windows. From the measurement of the coverage threshold in Fig. \ref{fig:BET} and from the parameters used to fit the BET curves, we can estimate the minimum temperature of the reservoir necessary to grow a light-induced film, at a given window temperature. The vapor pressure corresponding to the threshold is plotted in Fig. \ref{fig:threshold}(a) as a function of the window temperature, and the corresponding reservoir temperature threshold is plotted in Fig. \ref{fig:threshold}(b). 

For instance, for windows at $T_W=150$ $^\circ$C, the minimum temperature of the reservoir to observe the film growth is $T_R=115^\circ$ C. For $T_W=250^\circ$C the threshold of reservoir temperature is around $T_R=200^\circ$C. It is therefore understandable that the formation of the film is not commonly observed, as it requires relatively high temperatures of both the reservoir and the cell body. 

\bigskip

\includegraphics[width=8cm]{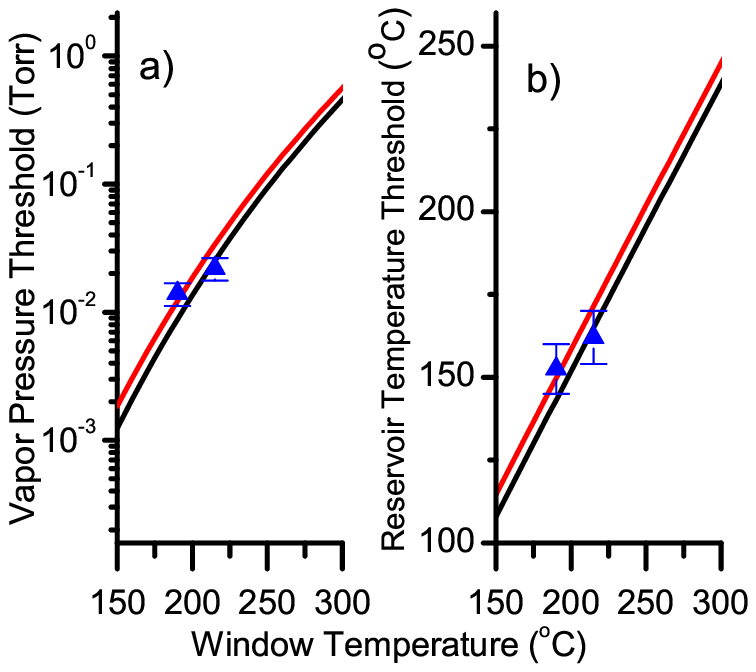}
\caption{(Color online). (a) Vapor-pressure threshold and (b) reservoir-temperature threshold for the growth of a light-induced film as a function of the temperature of the window. The lines were calculated using BET's model (Eq. (\ref{eq:BET})) and the parameters used to fit the curves shown in Fig. \ref{fig:BET}. For the black (solid) line we used the parameters of the curve with $T_W=215^\circ$ in Fig. \ref{fig:BET} while for the red (dashed) line we used the parameters of the curve with $T_W=190^\circ$ in Fig. \ref{fig:BET}. The interval between the two lines can be considered as an uncertainty due to imprecisions in the determination of the parameters used to fit BET's models in Fig. \ref{fig:BET}.}
\label{fig:threshold}
\end{figure}

\end{document}